\documentstyle[prb,aps,multicol,epsfig]{revtex}
\begin{document}

\title{Origin of the Optical Gap in Half-Doped Manganites }

\author {Mario Cuoco,$^1$ Canio Noce,$^1$
          and Andrzej M. Ole\'{s}$^{2,3}$ }
\address{$^1$I.N.F.M. -Unit\`a di Salerno, Dipartimento di
              Fisica "E.R. Caianiello", \\
              Universit\'a di Salerno, Via S. Allende,
              I-84081 Baronissi, Italy}
\address{$^2$Marian Smoluchowski Institute of Physics,
              Jagellonian University, Reymonta 4, PL-30059 Krak\'ow,
              Poland}
\address{$^3$Max-Planck-Institut f\"ur Festk\"orperforschung,
              Heisenbergstrasse 1, D-70569 Stuttgart, Germany}


\maketitle

\begin{abstract}
We have analyzed the coexisting charge and orbital ordering in
half-doped manganites using a model which includes Coulomb and
Jahn-Teller orbital polarization interactions. Most surprisingly,
the gap in the optical conductivity is reduced by both on-site and
intersite Coulomb interactions, but increases and explains the
experimental results when the Jahn-Teller terms with orbital
polarization are considered. The origin of this behavior is
explained within a molecular model which arises in the limit of
extreme topological frustration, when single electrons are
confined to molecular units consisting of three orbitals.
\end{abstract}
\pacs{PACS numbers: 75.30.Vn, 79.60.-i, 75.30.Et, 71.15-m}

\begin{multicols}{2} 

\section{Introduction}
\label{sec:intro}

The colossal magnetoresistance (CMR) manganites form a class of
materials in which the problem of the interplay between the spin,
orbital, and charge degrees of freedom is believed to lie at the
heart of understanding their physical
properties.\cite{Tok00,Ram97} The parent compound, LaMnO$_3$, is a
Mott insulator due to large local Coulomb interaction $U$, with a
single electron localized in degenerate $e_g$ orbitals at
Mn$^{3+}$ ions. The magnetic and orbital ordering follow then from
superexchange interactions which stabilize antiferromagnetic (AF)
order.\cite{Fei99} The doping by holes adds the charge degree of
freedom in systems like La$_{1-x}$Sr$_x$MnO$_3$, and triggers a
transition to ferromagnetic (FM) state, in which the kinetic
energy of the doped $e_g$ holes is optimized. This mechanism,
called double exchange, was postulated five decades ago by
Zener.\cite{Zen51} Yet, the AF order is favored again at half
doping ($x=0.5$),\cite{Ram97,Wol55} indicating that the AF
superexchange and the FM double exchange frustrate each other, as
in the FM metallic phase.\cite{Ole02}

The half-doped manganites are very peculiar, and exhibit a
so-called charge exchange (CE) AF ordering,\cite{Goo55,Wol55}
which involves FM zigzag chains that are staggered
antiferromagnetically; it has been observed experimentally by
neutron scattering \cite{Ste96,Kaw97} and by X-ray
diffraction,\cite{Mur98,Nak99} both in the one-plane
La$_{0.5}$Sr$_{1.5}$MnO$_4$, \cite{Ste96,Mur98} and in cubic
Nd$_{0.5}$Sr$_{0.5}$MnO$_3$ manganites.\cite{Kaw97,Nak99} The
mechanism of stability of the CE phase is still under debate. The
hopping between degenerate $e_g$ orbitals along zigzag chains
shows a {\it topological frustration\/} \cite{Hot00} which leads
to a band insulator.\cite{Sol99,Kho99} Although the CE phase is
then favored, the large on-site Coulomb element $U$ destabilizes
it for realistic parameters.\cite{She01} Therefore, one expects
that either intersite Coulomb interactions, \cite{Mut99,Jac00} or
the coupling to the lattice which leads to the Jahn-Teller (JT)
distortions,\cite{Miz97,Yun00} might play an important role.

We argue that, similar as for the FM metallic phase,\cite{Mac99}
the optical experiments may provide the clue to the proper
understanding of the CE phase within a truly relevant microscopic
model. The experiments for the La$_{0.5}$Sr$_{1.5}$MnO$_4$
compound \cite{Tok99,Jun00} show peculiar features by decreasing
temperature from 300 K to 10 K. Above the charge/orbital
transition $T_{CO}\simeq 217$ K, the midinfrared spectrum is broad
with an optical gap of the order of 0.2 eV and a width of about
1.5 eV. It centers around the main peak at $\sim 0.8$ eV. Below
$T_{CO}$ one observes a rigid shift toward high energy and a
narrowing of the spectrum. At the same time, the main peak shifts
towards the bottom of the spectrum, giving rise to an asymmetric
profile of the optical curve and an increase of the spectral
weight. Finally, as the temperature is lowered below the N\'eel
temperature $T_N\simeq 110$ K, a further slight and smooth
increase of the optical gap and narrowing of the main peak are
observed.

On the theoretical side, there have been only few reports on the
optical properties of the charge and orbital ordering in
half-doped manganites,\cite{Jun00,Sol01} based on the electronic
structure calculations. In fact, these studies can serve as a
starting point for the analysis of the correlated character of the
optical excitations in the case of $e_g$ electrons which are
coupled either via the Coulomb interaction, or via the JT
distortions.

In this paper we introduce a new microscopic model for half-doped
manganites. Next to the on-site Coulomb repulsion, it includes the
JT term responsible for the $e_g$ orbital polarization at
Mn$^{3+}$ ions, induced by the JT distortions around the occupied
sites when holes occupy the neighboring Mn$^{4+}$
ions.\cite{Wol55,Mah01} We study this model by exact
diagonalization (ED) using Lanczos algorithm,\cite{Jak00} and
investigate:
  (i) the microscopic origin of the CE phase, its stability and
      properties in presence of the JT terms,
 (ii) the nature of the optical excitations as due to
      transitions between the correlated states of $e_g$
      electrons at Mn($3d^4$) sites, and
(iii) the reasons for the peculiar changes observed in the optical
      spectra under increasing temperature.\cite{Tok99,Jun00}
The unexpected result of the reduction of the charge gap due to
electron correlations is explained as a consequence of the subtle
role played by the topological frustration. This behavior turns
out to be relevant in explaining the pseudogap like features
observed in the optical spectrum of the high temperature phase,
where the distortions are weak and the charge and orbital
correlations are only at short range. Furthermore, we show that
the large optical gap observed in La$_{0.5}$Sr$_{1.5}$MnO$_4$ at
low temperature \cite{Tok99} cannot be simply explained within
neither the electronic structure theory,\cite{Sol01} nor using a
correlated band insulator,\cite{Kho99} but instead by including
the JT interaction.

The paper is organized as follows. In Sec. II we present the model
Hamiltonian (Sec. II.A) and discuss the stability of the CE phase,
using as stabilizing mechanism either the JT term, or the
intersite Coulomb interaction (Sec. II.B). Next we elucidate the
frustration present in the kinetic energy due to the orbital phase
factors, and introduce a so-called {\it molecular model\/} (MM) in
which this frustration is maximized (Sec. II.C). The optical
spectra are determined and analyzed both for the CE phase and the
MM in Sec. III.A. The optical gap is extrapolated to the
thermodynamic limit using finite size scaling (Sec. III.B).
Finally, we discuss the obtained optical spectra in a broader
context of the existing experiments, and summarize our results in
Sec. IV.

\section{The model and its ground state}
\label{sec:model}

\subsection{Importance of the JT coupling}

The model we adopt follows from the experimental evidence in
manganites on the importance of the orbital degrees of freedom,
the local Coulomb interactions $\propto U$, and the JT distortions
$\propto E_{\rm JT}$. The local Coulomb repulsion $U$ reduces the
probability of double occupancies of $e_g$ orbitals, while the
Hund's exchange coupling $J_H$ between the $e_g$ and the $t_{2g}$
electrons is responsible for the high-spin states, realized both
at Mn$^{3+}$ and Mn$^{4+}$ ions. This term will not be explicitly
included, as we assume that the $e_g$ and $t_{2g}$ spins are
aligned, which corresponds to the frequently considered limit of
$J_H\to\infty$.

\begin{figure}
\includegraphics[width=6.2cm]{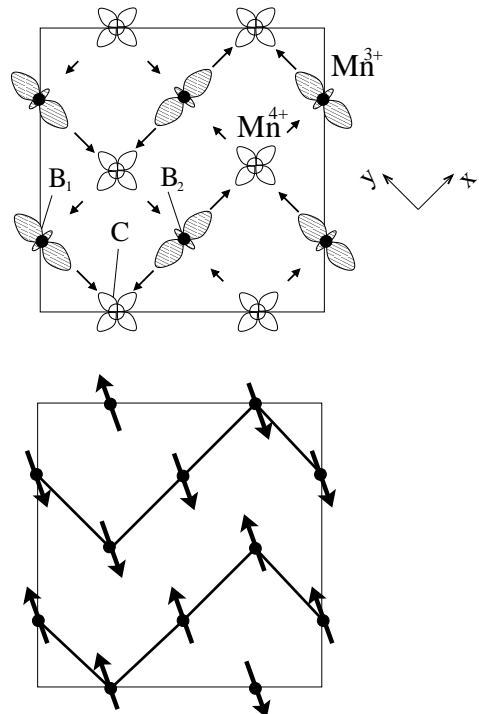}
\caption{ Schematic picture of the charge and orbital (top panel),
and of the spin configuration (bottom panel) for the CE phase. The
arrows in the upper panel indicate the in-plane oxygen distortions
around the manganese site at the bridge~(B$_1$,B$_2$) and
corner~(C) position, which induce the coupling between the $e_g$
electrons and the JT disortions $\{q_i\}$, as discussed in the
text. The length of the arrows in the top panel is related to the
amplitude of the distortion as due to the JT polaronic term.}
\label{fig:cefig}
\end{figure}

We consider (spinless) $e_g$ electrons moving along a FM zigzag
chain of the CE phase shown in Fig. \ref{fig:cefig}. Along each
chain two kinds of nonequivalent sites alternate: bridge ($i\in
B$) and corner ($j\in C$) sites (see Fig. \ref{fig:cefig}). The
{\it polaronic\/} model,
\begin{equation}
\label{model} {\cal H}_{\rm pol} = H_t + H_{U} + H_{\rm JT},
\end{equation}
consists of three terms,
\begin{eqnarray}
\label{Ht} H_t&=&\sum_{i\in B,j\in C}\Big[
             (-1)^{\lambda_{ij}}t_1(\phi)b_i^{\dagger}a_{jx}^{}\!-\!
                t_2(\phi)b_i^{\dagger}a_{jz}^{}\!+\!{\rm H.c.}\Big],
                                              \nonumber \\    \\
\label{Hu}
& &\hskip 1.2cm H_{U}= U\sum_{j\in C}n_{jx}n_{jz},            \\
\label{Hjt}
H_{\rm JT}&=&\sum_{i}\Big[g\sum_{j(i)}q_i(n_{i\zeta}-n_{i\xi})
                          (1-n_j)+\frac{1}{2}k_{JT}q_i^2\Big],
\end{eqnarray}
which stand for the kinetic energy ($H_t$), the on-site
(interorbital) Coulomb interaction $U$ at corner sites ($H_{U}$),
and the JT polaronic interaction ($H_{\rm JT}$), respectively. The
kinetic energy $H_t$ describes the hopping $\propto t$ of $e_g$
electrons between the directional $3x^2-r^2$ ($3y^2-r^2$) orbitals
at bridge ions, and the $x^2-y^2$ ($|x\rangle$) and $3z^2-r^2$
($|z\rangle$) orbitals at corner ions (see Fig. \ref{fig:cefig}),
with the corresponding electron creation operators:
$b_i^{\dagger}$, $a_{jx}^{\dagger}$, and $a_{jz}^{\dagger}$. The
hopping elements are: $t_1(\phi)=t\sin\phi$,
$t_2(\phi)=t\cos\phi$, and $\phi=\frac{\pi}{6}$, with $t$ standing
for the hopping between two identical directional orbitals along
the bond $\langle ij\rangle$, i.e., $3x^2-r^2$ orbitals along the
bond $\langle ij\rangle\parallel a$ axis. The phase factors
$(-1)^{\lambda_{ij}}$ follow from the alternating orbital phases
due to the zig-zag pattern, with $\lambda_{ij}=1$ or
$\lambda_{ij}=0$, where the hopping connects from the bridge
position $B_1$ or $B_2$ with the corner site $C$, respectively.
The JT orbital polarization interaction (\ref{Hjt}) splits off two
$e_g$ orbitals at site $i$, if a neighboring site $j(i)$ is
occupied by a hole. This interaction has a similar form to that
deduced for orbital polarons \cite{Kil99} --- it favors the
occupancy of the directional $|\zeta\rangle$ orbital, oriented
along the bond $\langle ij\rangle$, over the orthogonal to it
planar $|\xi\rangle$ orbital (e.g., $|3x^2-r^2\rangle$ over
$|y^2-z^2\rangle$ orbital for a bond along $a$ axis). The
classical (phonon) variable $q_i$ stands for the respective oxygen
distortions around an electron at site $i$, either a bridge or a
corner position along the chain (Fig. \ref{fig:cefig}), while $g$
is the coupling constant between $e_g$ electrons and distortions of
MnO$_6$ octahedra, and $k_{JT}$ is the spring constant for the JT
mode distortions. It is convenient to introduce the JT energy,
defined by $E_{JT}=g^2/k_{JT}$, which is naturally given by the
scaling of ${q}_i=(g/k_{JT})\tilde{q}_i$, where $g/k_{JT}$ is the
typical length scale for the JT distortion. By means of these
substitutions, $H_{\rm JT}$ reads as:
\begin{eqnarray}
\label{Hjt1}
H_{\rm JT}=E_{JT}\sum_{i}\Big[\sum_{j(i)}\tilde{q}_i(n_{i\zeta}
          -n_{i\xi})(1-n_j)+\frac{1}{2} \tilde{q}_i^2 \Big].
\end{eqnarray}

The JT term couples corner and bridge sites in an $(a,b)$ plane,
either along the considered zigzag chain, or between two adjacent
chains. We minimized the total energy $\langle {\cal H}_{\rm
pol}\rangle$ to eliminate the phonon variables $\{\tilde{q}_i\}$,
which gives the electronic Hamiltonian solved selfconsistently
with:
\begin{equation}
\label{qcla}
\tilde{q}_i=-\sum_{j(i)}
             \langle (n_{i\zeta}-n_{i\xi})(1-n_j)\rangle.
\end{equation}
The classical variables at bridge and corner positions are
different, as suggested by different oxygen distortions around Mn
ions.\cite{Mah01} If one considers the form of $\tilde{q}_i$, it
is easy to recognize that the distortions are weak within this
approximation, if the populations of the two orbitals are the
same. Moreover, even in presence of a homogeneous orbital
distribution, a small distortion pattern might be induced by the
nearest neighbor orbital correlations.
\begin{figure}
\includegraphics[width=6.5cm]{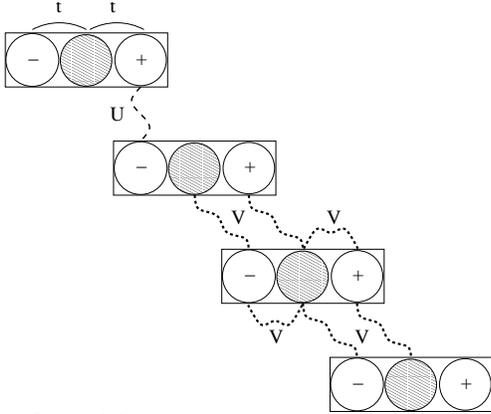}%
\caption{ Schematic representation of the orbital structure in the
MM. The chain is divided into molecular triads constituted of a
bridge configuration (gray circle) together with an odd (circle
with $-$) and even (circle with $+$) superposition of the corner
orbitals. The solid line shows the hopping processes inside the
molecule, while the dashed lines indicate the intra- and
inter-molecular interactions which result from the $U$ and $V$
coupling.} \label{fig:molmod}
\end{figure}
We would like to point out that the form of the interaction
between the JT modes and the $e_g$ electrons is different from
that usually assumed, where the on-site $Q_2$ and $Q_3$ mode are
coupled to the local orbital density.\cite{Yun00} As regards the
present model, we are assuming that a directional JT ($Q_3$ like)
mode, with its direction selected by the considered bond $\langle
ij\rangle$, is activated by the presence of a hole on the neighbor
manganese, thus inducing a non-local correlated dynamics between
the lattice and the charge degrees of freedom. As a direct
consequence of this non-local coupling, one ends up with an
effective directional Coulomb interaction between nearest neighbor
manganese which strongly influences both the ground state and the
properties of the low-energy excitations.

\subsection{Stability of the CE phase}

\begin{figure}
\includegraphics[width=7.cm]{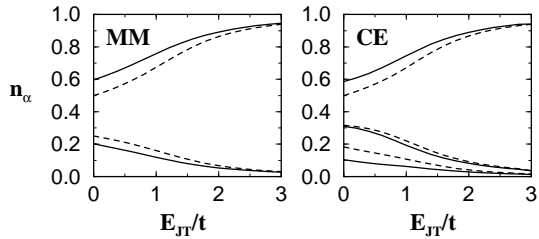}
\caption{ Electron densities $n_{\alpha}$: $n_b$, $n_x$ and $n_z$
from top to bottom, as functions of $E_{JT}$, obtained using ED
for the chain of $L=12$ atoms at $U=0$ (dashed lines) and at
$U=10t$ (solid lines) in: MM ($\phi=\frac{\pi}{4}$), and
                  CE phase ($\phi=\frac{\pi}{6}$).}
\label{fig:n}
\end{figure}

We begin with the investigation of stability of the CE phase with
respect to the C phase, which consists of linear FM chains along
the $b$ axis staggered in the $a$ direction, using the mean-field
(MF) approximation. The C phase (01) is described by ${\cal
H}_{\rm pol}$ (\ref{model}) at $\phi=\frac{\pi}{2}$, with the
hopping element $t$ between directional $3y^2-r^2$ orbitals along
each FM chain. While the on-site Coulomb element $U$ increases the
energy of the CE phase, it plays no role in the C phase, as the
$x^2-z^2$ orbitals are empty. Thus, the C phase is favored above
$U\simeq 2.7t$ ($U\simeq 12.0t$) in the MF (ED) method, if $E_{\rm
JT}=0$. We have verified that increasing $E_{\rm JT}$ induces in
MF a charge disproportionation $\delta n$ in both phases, but this
effect is more pronounced in the CE phase, lowering its energy
especially when $E_{\rm JT}>t$. Thus, we recognize that the JT
term ${\cal H}_{\rm JT}$ in Eq. (\ref{model}) is the main
stabilizing mechanism of the observed CE phase.\cite{notej} The JT
term induces a charge ordering (coexisting with orbital ordering),
which is in principle unlimited and may come close to $\delta
n=1$, unlike the weak charge ordering induced by the Coulomb
interaction $U$, for which the order parameter cannot exceed
$\delta n=0.19$.\cite{Kho99} Already at $U=0$ the charge order is
more pronounced than that induced by $U\to\infty$ in a broad range
of $E_{\rm JT}>0.7t$ (Fig. \ref{fig:n}). At finite $E_{\rm JT}$ the
charge disproportionation is only weakly enhanced by the on-site
Coulomb element $U$. Although the form of the JT interaction term
is different, our results agree qualitatively with the recent {\it
ab initio\/} density-functional calculations.\cite{Pop02}

\begin{table}
\caption{Eigenstates and eigenenergies for a molecular unit $i$ in
the MM ($\phi=\frac{\pi}{4}$) at $U=E_{\rm JT}=0$. The bonding,
nonbonding, and antibonding states are labelled as $|B\rangle$,
$|N\rangle$, and $|A\rangle$.} \vskip .2cm
\begin{tabular}{ccllc}
& state    & creation operator & energy  \cr \tableline &
$|B\rangle$ & $B_i^{\dagger}=\frac{1}{\sqrt{2}}b_i^{\dagger}
   +\frac{1}{2}(a_{i-1,+}^{\dagger}+a_{i+1,-}^{\dagger})$  &
   $E_{B}=-\sqrt{2}t$ & \cr
& $|N\rangle$ & $N_i^{\dagger}=\frac{1}{\sqrt{2}}
   (a_{i-1,+}^{\dagger}-a_{i+1,-}^{\dagger})$ &  $E_{N}=0$  & \cr
& $|A\rangle$ & $A_i^{\dagger}=\frac{1}{\sqrt{2}}b_i^{\dagger}
   -\frac{1}{2}(a_{i-1,+}^{\dagger}+a_{i+1,-}^{\dagger})$  &
   $E_{A}=\sqrt{2}t$  & \cr
\end{tabular}
\label{tab:mass}
\end{table}
The CE phase ground state could also be reproduced by intersite
Coulomb interactions in the {\it electronic\/} model,\cite{Jac00}
\begin{equation}
\label{modele} {\cal H}_{\rm elec} = H_t + H_{U} + H_{V},
\end{equation}
with $H_{V}=V\sum_{\langle ij\rangle}n_{i}n_{j}$. Taking
$V\simeq\frac {1}{2}E_{\rm JT}$ the models (\ref{model}) and
(\ref{modele}) lead indeed to similar MF phase diagrams. This
might suggest that ${\cal H}_{\rm elec}$ could also provide a
valid explanation for the observed CE phase. However, going {\it
beyond\/} the MF one finds that the two models are fundamentally
different and lead to qualitatively different results, both for
the ground state,\cite{notegs} and for the excitation spectra.

\subsection{Molecular model}

Perhaps the most peculiar feature of the CE phase is the topological
frustration of the kinetic energy which occurs due to conflicting
orbital phases,\cite{Hot00} and leads to a band insulator already
at $U=E_{\rm JT}=0$.\cite{Kil99} We emphasize that this frustration
is maximized when the hopping elements with the opposite phases
precisely compensate each other, i.e., $t_1=t_2=\frac{1}{2}t$, in
the MM at $\phi=\frac{\pi}{4}$ (see Fig. \ref{fig:molmod}).
Transforming then the orbital basis
$\{a_{ix}^{\dagger},a_{iz}^{\dagger}\}$ to even and odd symmetry
states at each corner site:
\begin{equation}
\label{plusminus}
a_{i\pm}^{\dagger}=\frac{1}{\sqrt{2}}(a_{iz}^{\dagger}\pm
                                      a_{ix}^{\dagger}),
\end{equation}
one finds that the hopping term simplifies to:
\begin{equation}
\label{htmm}
H_t=t\sum_{i=2n}\Big(
                b_i^{\dagger}a_{i-1,+}^{}
               +b_i^{\dagger}a_{i+1,-}^{}+{\rm H.c.}\Big),
\end{equation}
and represents a superposition of molecular units. Each of them
consists of a central ($i=2n$) bridge orbital $b_i^{\dagger}$, an
even symmetry $a_{i-1,+}^{\dagger}$ orbital at the preceding site
$i-1$, and an odd symmetry $a_{i+1,-}^{\dagger}$ orbital at the
following site $i+1$, with the eigenstates given in Table I.

In the ground state at half doping,
\begin{equation}
\label{gsmm}
|\Phi_0\rangle=\prod_i B_i^{\dagger}|0\rangle,
\end{equation}
all $e_g$ electrons are confined to individual molecular units and
occupy the bonding states, and the energy per site $E_{\rm MM}=
-0.707t$ is significantly lower than in the CE phase, $E_{\rm CE}=
-0.695t$. It is interesting to realize that although the electrons
are now partly localized within the molecular units, the kinetic
energy is at the minimum allowed in the subspace of
$t_1^2+t_2^2=t^2$.

While the density distribution $\{n_{\alpha}\}$ ($\alpha=b,x,z$)
is uniform for the noninteracting electrons, with the same density
at bridge ($n_b=n_x+n_z$) and corner ($n_c$) positions, increasing
$U$ or $E_{\rm JT}$ polarizes the system and induces a charge
transfer towards bridge atoms. The effect of $U$ is somewhat more
pronounced in the MM than in the CE phase. This is best
illustrated by considering the Gutzwiller projected state at
$U=\infty$, \cite{Kho99}  which leads to the kinetic energy:
\begin{equation}
\label{mm}
{\tilde H}_t=t\sum_{i=2n}\Big(
             b_i^{\dagger}{\tilde a}_{i-1,+}^{}
            +b_i^{\dagger}{\tilde a}_{i+1,-}^{}+{\rm H.c.}\Big),
\end{equation}
where ${\tilde a}_{j,\pm}^{}=a_{j,\pm}^{}(1-n_{j,\mp}^{})$ are the
electron annihilation operators in the restricted space which does
not contain double occupancies. By decoupling the quartic terms in
${\tilde H}_t$,\cite{Kho99} $b_i^{\dagger}a_{j,\pm}^{}n_{j,\mp}^{}
\simeq b_i^{\dagger}a_{j,\pm}^{}\langle
n_{j,\mp}^{}\rangle+\langle b_i^{\dagger}a_{j,\pm}^{}\rangle
n_{j,\mp}^{}$, we have found a somewhat larger charge anisotropy
$\delta n=n_b-(n_x+n_z)\simeq 0.28$ for the MM than 0.19 found
before in the CE phase.\cite{Kho99} A much stronger charge
anisotropy, however, is induced in both cases by the JT term; with
increasing $E_{\rm JT}$ it approaches gradually the localized
limit, and at $E_{\rm JT}=2t$ one finds already $\delta n>0.65$
(Fig. \ref{fig:n}), with $n_x>n_z$ due to the different hopping
elements in the CE phase. The on-site Coulomb repulsion (\ref{Hu})
is invariant under the transformation to $\{|+\rangle,
|-\rangle\}$ orbitals and gives the interaction $Un_{i-}n_{j+}$
between two electrons from adjacent molecular units in the MM. It
is quite remarkable that the effect of $U$ in the ground state is
weak in the regime of $E_{\rm JT}>t$ (Fig. \ref{fig:n}). This
demonstrates that the correlation effects are {\it partly
suppressed\/} by the coupling to the lattice, similar as for the
on-site JT term.\cite{Yun00}

\section{The optical conductivity}
\label{sec:optic}

We now turn to the optical excitations and analyze the evolution
of the optical gap $\Delta$ with the increasing Coulomb interactions
and the JT coupling. The charge gap has been extracted by the
calculation of the optical conductivity $\sigma(\omega)$ within a
generalization of the Lanczos diagonalization
technique.\cite{Jak00} Since the system is in an insulating state,
one can restrict the calculation of $\sigma(\omega)$ to the finite
frequency response given by the Kubo formula
\begin{eqnarray}
\label{sigma}
\sigma(\omega)=\frac{1-e^{-\beta \omega}}{\omega} Re
\int_0^{\infty} d\tau\;e^{i \omega \tau} \langle j(\tau)j\rangle,
\end{eqnarray}
where $\beta=1/k_{B}T$ and $j$ stands for the contribution of the
current operator along the FM zig-zag chain, obtained by the
projection at the $x(y)$ direction:
\begin{equation}
\label{j}
j=i \sum_{\langle lm\rangle,\alpha} t_{lm}R_{lm}
({b}^{\dagger}_la_{m\alpha}^{}-{a}_{m\alpha}^{\dagger}b^{\dagger}_l).
\end{equation}
Here $R_{lm}$ is the component of the vector which connects
nearest neighbor sites $l$ and $m$, $t_{lm}$ is the hopping
element between bridge and corner sites as defined above, and
$\alpha=(x,z)$, respectively. By applying an electric field along
the zig-zag chain, shown in Fig. \ref{fig:cefig}, one accelerates
charge along the chain, either along the $x$ or along $y$ axis,
depending on the bond direction. In this way the current includes
the charge-transfer processes from $B_1$ or $B_2$ to $C$ sites,
and {\it vice versa\/}.

\subsection{ Interaction effects in the optical spectra }

\begin{figure}
\includegraphics[width=6.5cm]{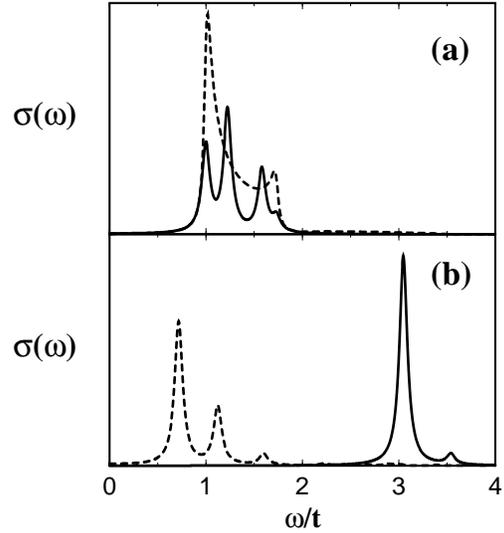}%
\caption{ Optical spectra $\sigma(\omega)$ as obtained in ED for
the chain of $L=12$ atoms in the polaronic model (\ref{model})
with $\phi=\frac{\pi}{6}$, and for: (a) $U=E_{JT}=0$, ED result
(full line) compared with an exact
    spectrum for CE phase (dashed line);
(b) $U=10t$, $E_{JT}=0$ and $E_{JT}=1.8t$ (dashed and full line).
} \label{fig:spectra}
\end{figure}

\begin{figure}
\includegraphics[width=7.cm]{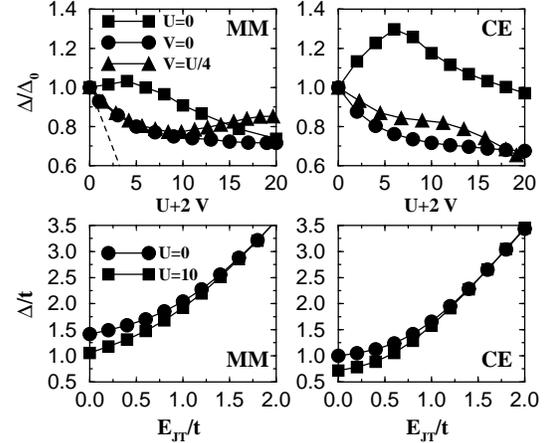}%
\caption{ Optical gaps $\Delta$ found in ED for the MM (left) and
the CE phase (right) as functions of electron-electron
interactions (top): -- $U$ (circles), $V$ (squares), and $U+2V$
for $V=U/4$ (triangles), and of $E_{JT}$ (bottom) for: $U=0$
(circles) and $U=10t$ (squares). Dashed line in top left panel
shows the weak coupling expansion [Eq. (\protect\ref{n+})]. }
\label{fig:gap}
\end{figure}

\begin{figure}
\includegraphics[width=7.cm]{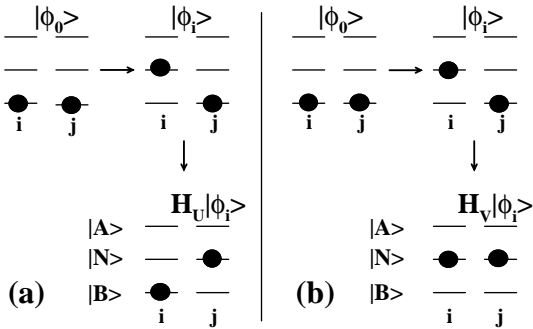}%
\caption{ Schematic representation of the low-energy optical
excitations induced in the MM by: (a) on-site $U$, and (b)
intersite $V$ Coulomb interaction. The lowest-energy excitation
$|\phi_i\rangle$ is made starting from the ground state
$|\phi_0\rangle$. The terms $\propto U$ permute the bonding
$|B\rangle$ and non-bonding $|N\rangle$ states on neighboring
molecules and allow a propagation of the low-energy excitation
along the chain (a), while the terms $\propto V$ provide only the
coupling of the low-energy optical excitations to higher energy
configurations which contain more than one nonbonding state in the
chain (b).} \label{fig:toy}
\end{figure}

First, we investigate the charge transport along the chain by
analyzing the position of the lowest energy excitation in the
optical spectra, which allows to extract the dependence of the
optical gap on the value of the JT coupling constant and the
electron-electron interaction parameters. Noninteracting electrons
give a band structure gap $\Delta_0=t$ in the CE phase (shown in
Fig. \ref{fig:spectra}),\cite{Sol01} and $\Delta_0=\sqrt{2}t$ in
the MM. The on-site repulsion $U$ typically enhances the energy of
local excitations in the correlated systems, and increases the
optical gap.\cite{Esk94} Thus, it was unexpected that the gap first
{\it decreases\/} linearly with $U$ at $E_{\rm JT}=0$, and next
{\it saturates\/} at finite $\Delta\simeq 0.7\Delta_0$ in the regime
of $U\gg t$, both in the CE phase and in the MM (Fig. \ref{fig:gap}).
This behavior can be understood, however, by considering the
lowest optical excitation in the MM for $U\simeq t$,
$|\Phi_i\rangle=N_i^{\dagger}B_i^{} |\Phi_0\rangle$, and expanding
the interaction term $\propto Un_{i-}n_{j+}$ using the Hubbard
operators ($X_j^{AB}=A_j^{\dagger}B_j^{}$, {\it etcetera}),
\begin{eqnarray}
\label{n+} n_{i\mp1,\pm} &=&
\frac{1}{4}[X_i^{BB}+X_i^{AA}+2X_i^{NN}-X_i^{AB}-X_i^{BA} \nonumber \\
& &      \mp\sqrt{2}(X_i^{AN}+ X_i^{NA}-X_i^{BN}-X_i^{NB})].
\end{eqnarray}
Indeed, the diagonal term $\propto X_i^{NN}$ enhances the
excitation energy by $\frac{1}{8}U$, suggesting that the gap
$\Delta$ should increase. However, in spite of the electron
localization within the molecular units, the above excitation is
not localized, but {\it can propagate\/} along the chain due to
the terms $\propto X_i^{BN}X_{i+2}^{NB}$ in Eq. (\ref{n+}) [see
Fig. \ref{fig:toy}(a)] with dispersion $-\frac{1}{4}U\cos k$,
resulting in a {\it decrease\/} of the optical gap calculated at
band minimum, $\Delta\simeq\Delta_0-\frac{1}{8}U$.

A similar weak coupling expansion performed for the intersite
Coulomb term $\propto V$ in Eq. (\ref{modele}) gives rise in
lowest order to the same optical excited state
$\{|\Phi_i\rangle\}$, but it cannot propagate as now ${\cal
H}_{\rm elec}$ does not contain terms of the form $\propto
X_i^{BN}X_{i+2}^{NB}$. Therefore, acting on such states by ${\cal
H}_{\rm elec}$ leads only to higher energy local excitations [Fig.
\ref{fig:toy}(b)], and explains why the optical gap increases
first in the range of $2V\lesssim 5t$, before it starts to
decrease at higher values of $V$ (Fig. \ref{fig:gap}). However, in
the realistic regime of parameters ($V\ll U$ and $V\propto U$) the
gap would decrease, as dominated by the on-site term $U$. The gaps
in the CE phase follow the same trends when the electron
interactions are increased (Fig. \ref{fig:gap}), with the optical
gap being reduced to $\Delta\simeq 0.7t$ at $U=10t$ (Fig.
\ref{fig:spectra}). Thus, the weak propagation along the zigzag
chain at $\phi=\frac{\pi}{6}$ does not change the generic picture
of the gap decreasing with $U$, which emerges in the MM.

In contrast, the interactions with the lattice described by the JT
term lead to qualitatively different physics. Increasing $E_{\rm
JT}$ decreases local potentials at bridge positions in the MM, and
thus the splitting between the $|B\rangle$ and $|N\rangle$ state
and the gap value $\Delta$ {\it increase\/}. Although the gaps in
the MM and CE phase are different in the realistic regime of
$E_{\rm JT}\simeq t$, they approach the same limit $\Delta \simeq
2E_{\rm JT}$ at large $E_{\rm JT}\gg t$. Furthermore, even if the
JT term contains an effective long-range Coulomb term mediated by
the JT distortions, this contribution does not counterbalance the
increase of the charge gap due to the splitting between the bridge
and corner states.
We also investigated the changes in the shape of the optical
spectra $\sigma(\omega)$ with increasing interactions. First of
all, the asymmetric spectrum for noninteracting electrons, with a
large spectral weight at the low-energy feature, is well
reproduced (Fig. \ref{fig:spectra}). Both increasing $U$ and
$E_{\rm JT}$ lead to more pronounced asymmetry of the spectra. By
analyzing the spectral weight distribution, we established that
electrons are excited from bridge (Mn$^{3+}$) to corner
(Mn$^{4+}$) sites, and the $x$ ($z$) character of the excitations
is found predominantly close to the low (high) energy edge. The
high energy part looses its spectral weight faster, and the
distance between these two features decreases with increasing
$E_{\rm JT}$ or $U$, resulting in much narrower optical spectra
[e.g., the distance between two local excitations described by Eq.
(\ref{mm}) is only $0.87t$ at $U=\infty$ compared with $1.41t$ at
$U=0$].

\subsection{ Finite size scaling }

\begin{figure}
\includegraphics[width=7.cm]{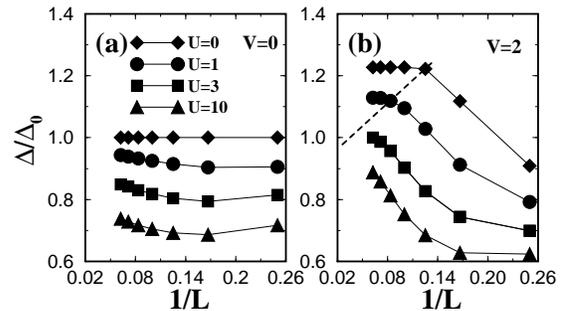}%
\caption{ Optical gap $\Delta$ in the CE phase (in the units of
the gap for noninteracting electrons $\Delta_0$), found by means
of Lanczos diagonalization for different chain size
$L=4,\cdots,16$, for $V=2t$ and: $U=0$, 1, 3, $10t$. The dashed
line in (b) indicates the linear interpolation for the critical
size above which the charge gap saturates.} \label{fig:gapscaled}
\end{figure}

In order to extract information about the electronic system in the
thermodynamic limit, we have performed a scaling analysis of the
charge gap obtained for different values of the chain lengths (from
$L=4$ to $L=16$ sites), and two representative sets of parameters
with $E_{JT}=0$: $U=0$, 1, 3, and $10t$, taking either $V=0$ or
$V=2t$. We begin with the analysis of the influence of increasing
$U$ at $V=0$ [Fig. \ref{fig:gapscaled}(a)]. In this case the charge
gap is almost independent on the cluster size, which indicates that
on-site excitations discussed in the previous Section indeed
reduce the optical gap due to correlation effects.

To understand the results at $V=2t$, let consider first the case
of $U=0$. As shown in the Fig. \ref{fig:gapscaled}(b), the behavior
of the charge gap is monotonic up to $L=8$ and then it saturates.
One can observe that the only effect of $V$ is to confine the
charge in the ground state as the lowest charge excitation has a
gap in the spectrum, and in addition to localize the lowest
optical excitation on a length which is smaller than $L=8$, as the
charge gap does not change above this critical size. In other
words, the results suggest that the quantum coherence of the lowest
optical excitation extends only over a limited distance, so that
when the cluster size exceeds $L=8$ length, the increasing further
cluster size does not affect the value of the charge gap anymore.

Consider now the effect of the on-site interaction $\propto U$.
As one can see in the Fig. \ref{fig:gapscaled}, increasing $U$
reduces the amplitude of the charge gap, and modifies the critical
size, above which the gap saturates. Due to the limitation in the
numerical computation, we can follow the evolution of the critical
size (dashed line) only for $U\simeq t$. Indeed, for large values
of $U\gg t$, one observes only a monotonic growth of $\Delta$ as a
function of $1/L$, without any sign of the gap saturation. We have
extracted the behavior of the critical size vs $U$ by means of a
linear fit. One can argue that the effect of $U$ is to increase
the localization length in the local optical excitation spectrum
(LOES), which yields a larger value of chain size to get a
saturation in the charge gap. This itinerant character induced by
$U$ is clear in the MM, where we have seen that finite $U$ allows
for a propagation of the LOES.

By summarizing the results of the scaling, we deduced that the
effect of $V$ is to harden the charge fluctuations in the ground
state and to induce a localization of the LOES on a length scale of
about $L=8$. On the contrary, the $U$ term induces an itinerancy in
the LOES which manifests itself by an increase of the localization
length in the LOES, and consequently of the critical size of
saturation of the charge gap. The competition of these two
counteracting effects turns out to give the reduction of the charge
gap in the bulk limit only for a very weak intersite Coulomb
repulsion, $V\ll U$.

A further important conclusion from the presented analysis is that
one finds a large range of variation of $\Delta/\Delta_0$, when
one moves from a chain of $L=4$ sites to the extrapolated bulk
limit. This analysis might be relevant for real materials if one
interprets the disordered high temperature phase of the half-doped
manganites as made by a liquid of zig-zag FM chains with different
length, so that the optical response has to be built by making a
superposition of the response of chains with any size. We argue
that the low-energy part of the spectra contains this fluctuating
superposition of several chain contributions of different size,
and these short-chain contributions are suppressed when one enters
into the ordered phase below $T_{CO}$.

\section{Discussion and summary}
\label{sec:summa}

We argue that the present model agrees qualitatively with
available experimental information, and our findings reproduce the
experimental observations in La$_{0.5}$Sr$_{1.5}$MnO$_4$
\cite{Tok99} and Nd$_{0.5}$Ca$_{1.5}$MnO$_3$.\cite{Oki02} In the
high temperature regime when the JT distortions are weak, the
optical gap is small and dominated by the effect of electronic
correlations. When the temperature is decreased below $T_{CO}$,
the spectral weight grows and shifts to higher energy, the optical
spectrum is narrowed, and the asymmetry in the spectrum increases.
All these features {\it can be explained\/} within the present
polaronic model (\ref{model}) as an indication of the onset of the
orbital and charge ordering from the disordered state, described
by finite values of local variables $\{\tilde{q}_i\}$, and thus
increasing orbital polarization due to the JT terms $\propto
E_{\rm JT}\tilde{q}_i$. Further shift of the peak feature up to
$\Delta\simeq 1.3$ eV would follow from the increased
$\{\tilde{q}_i\}$ below the magnetic transition, when the magnetic
fluctuations are suppressed. The observed change of the gap
$\Delta$ is reproduced by $E_{\rm JT}\simeq 1.2t$, which
corresponds with $t=0.4$ eV \cite{Fei99} to the expected orbital
splitting of $\sim 0.3$ eV.\cite{Kil99}

It is worth pointing out that previous studies have discussed the
optical conductivity in the CE phase as due to dipole transitions
from Mn($e_g$) to Mn($4p$) states by means of tight-binding
approaches and first principle band-structure calculations in the
local spin-density approximation.\cite{Sol01} The analysis of Ref.
\onlinecite{Sol01} included the effect of the JT distortion as an
effective energy level splitting between the bridge and corner
states. The proposed interpretation that the changes of the
optical spectra are mainly due to the JT distortion on the
Mn($e_g$)--Mn($4p$) dipole transitions does not include the effect
of correlations, and thus it fails to reproduce the experimentally
observed modification of the spectral weight when the temperature
is lowered.\cite{Tok99}

Our interpretation of the optical spectra is different. We have
shown that the low-energy optical excitations are mainly due to
the interband transitions between correlated Mn($e_g$) states. The
effect of the JT coupling $\propto E_{JT}$ may be summarized as
follows:
  (i) the uncorrelated spectrum shifts toward higher energy,
 (ii) the spectral weight increases in the low-energy part, and
(iii) the asymmetry of the profile grows as the spectral weight in
      the high energy shoulder gets suppressed.
Particularly the second feature is strongly dependent on the form
of the JT coupling. Indeed, by a simple level splitting of the
bridge and corner states, one would just obtain a rigid shift of
the spectrum, while the effective directional Coulomb interaction
present in H$_{JT}$ (\ref{Hjt1}) is responsible of the increase of
spectral weight in the low energy side.

Finally, our results show that the shift of the main peak and the
increase of its spectral weight come essentially as a consequence
of the orbital polarization present in the JT coupling term. It
influences the optical transitions between correlated $e_g$
states. The pseudogap activation in the high temperature phase is
mainly due to the effect of Coulomb correlation which reduces the
band gap. Moreover, as mentioned above, the low energy tail in the
spectra observed in the high temperature regime, might originate
from the superposition of the optical response of FM zig-zag chains
with different length (see Fig. \ref{fig:gapscaled}).

In summary, we have shown that the subtle interplay between the
topology, the Coulomb correlations and the orbital polarization JT
terms are responsible for the temperature dependent optical
properties of La$_{0.5}$Sr$_{1.5}$MnO$_4$. Thus, we argue that
these interactions play a crucial role for the observed coexisting
orbital and charge ordering in half-doped manganites, which cannot
be explained within purely electronic models. When the orbital
polarization interaction is taken into account, the charge
disproportionation turns out to be much stronger than that induced
by $U$ alone, but not as pronounced as in the original Goodenough
picture.

\acknowledgments
We thank P. Horsch and Y. Tokura for valuable discussions. This
work was supported by the Committee of Scientific Research (KBN)
Project No. 5~P03B~055~20.



\end{multicols}


\begin{references}
\bibitem{notej} The CE phase is further stabilized by the AF
                superexchange terms. While the $t_{2g}$ superexchange
                is exactly the same in both phases, the stability of
                the CE phase is further improved by the $e_g$
                superexchange in the realistic regime of $U\simeq
                10t$.\protect\cite{Fei99,Ole02}. These terms
                distinguish between different orientation of the
                occupied $e_g$ orbitals at Mn$^{3+}$ ions and favor
                the CE phase.
\bibitem{notegs} Taking $U=10t$ which corresponds to the high-spin
                charge excitation (Ref. \protect\onlinecite{Fei99}),
                one finds that the CE phase has still a higher energy
                than the C phase at $V<U/2$ using ${\cal H}_{\rm elec}$.
                In contrast, the local distortions $\{q_i\}$ which
                simulate the oxygen distortions are much weaker in the
                C phase than in the CE phase when ${\cal H}_{\rm pol}$
                is used, and thus the CE phase is then more stable for
                the same value of $U=10t$.
\end{references}
\end{document}